\documentstyle[11pt,newpasp,twoside,epsf]{article}
\begin{document}
\title{Studying Galaxy Formation with Hubble's Successor}
\author{Rachel S. Somerville}
\affil{Department of Astronomy, University of Michigan, Ann Arbor, MI 48109}

\begin{abstract}
In this paper, I discuss the capabilities and limitations of an 8-10\,m
ultraviolet/optical telescope in space, the proposed successor to the Hubble
Space Telescope,
in the context of galaxy studies.  The exquisite spatial resolution
and excellent sensitivity of such a facility would open up new
possibilities for the study of nearby dwarf galaxies ($z \la 0.5$),
and for studying the internal structure and kinematics of more
luminous galaxies at high redshift ($z \ga 2$). These applications are
of particular importance because they would address areas in which the
popular Cold Dark Matter  theory is in potential conflict with
observations.
\end{abstract}

\section{What Do We Need to Know About Galaxy Formation?}
The ``Cold Dark Matter'' (CDM) or hierarchical paradigm of structure
formation provides a useful framework for attempting to understand
galaxies and cosmology.  As the values of the cosmological parameters
within this framework have become more and more tightly constrained by
non-galaxy-based observations (like the cosmic microwave
background and supernovae), most of
the major uncertainties have to do with the messy ``gastrophysics''
that connects dark matter with the gas and stars that we can observe
directly. Some of these issues include:

\begin{itemize}
\item {\bf Cooling} --- observations indicate that less of the gas in
the Universe has cooled than is predicted by simulations. In
particular, cooling in clusters seems to have been ``shut off'' by
some unknown process.

\item {\bf Star formation} --- what determines how efficiently a
galaxy can turn cold gas into stars? How does this efficiency scale
with redshift and galaxy properties? What is the physical basis of
empirical scaling laws such as the ``Kennicutt Law'' (Kennicutt 1989;
1998), and are they universal? What factors determine the duration and
efficiency of the bursts of star formation seen in interacting
galaxies? Is the stellar initial mass function universal in space and
time, and why does it have its observed shape?

\item {\bf Stellar feedback} --- how does the energy from massive
stars and supernovae affect the interstellar medium (ISM), intergalactic
medium (IGM), and intracluster medium (ICM), and how does this
impact future generations of stars?  Is the main effect predominantly
thermal or kinetic?  How important are turbulence and large-scale
galactic winds and outflows?

\item {\bf Heavy elements and dust} --- how efficiently are metals
expelled from the potential wells of galaxies? How did the ICM and IGM
get so uniformly polluted with heavy elements?  Is there a universal
dust-to-metal ratio in all galaxies?  What determines the degree of
optical/ultraviolet extinction a galaxy experiences? Do the properties of dust
in galaxies (composition, temperature, etc.) differ dramatically from
one galaxy to another, or as a function of redshift? 
\end{itemize}

While the observations obtained with the new generation of space-based
facilities and the large number of 6-10\,m-class ground-based telescopes
now coming on line will doubtless bring us a much better understanding
of many of these issues, I expect that even 10--15 years from now,
some of them will not be completely put to rest. In the remainder of
this paper, I focus on a few specific examples of observations that
would utilize the unique capabilities of the proposed Hubble Space 
Telescope successor (the ``Next HST'' -- NHST) to address
important theoretical questions about galaxy formation.

\section{What Could We Do with a 10\,m Optical Telescope in Space?}
At the risk of stating the obvious, I shall go through some of the
capabilities and limitations of the proposed NHST facility for
studying galaxies.  The ultraviolet/optical wavelength sensitivity is optimal
for studying objects with young stellar populations. There are many
indicators of star-formation activity in this wavelength range,
although they are suspect due to the uncertain effects of dust.
Because of the large aperture and the location in space, NHST would
have exquisite spatial resolution. Space-based observations also have
lower background and so lower flux limits. This is useful for looking
at high redshift or intrinsically faint and/or low surface brightness
objects (a familiar lesson from HST). However, by $z\simeq6$, the
Lyman break has redshifted into the I-band, so a purely optical
telescope will not be useful for observing galaxies at $z\ga6$ or so
(but that is what NGST is for).

\begin{figure}
\plottwo{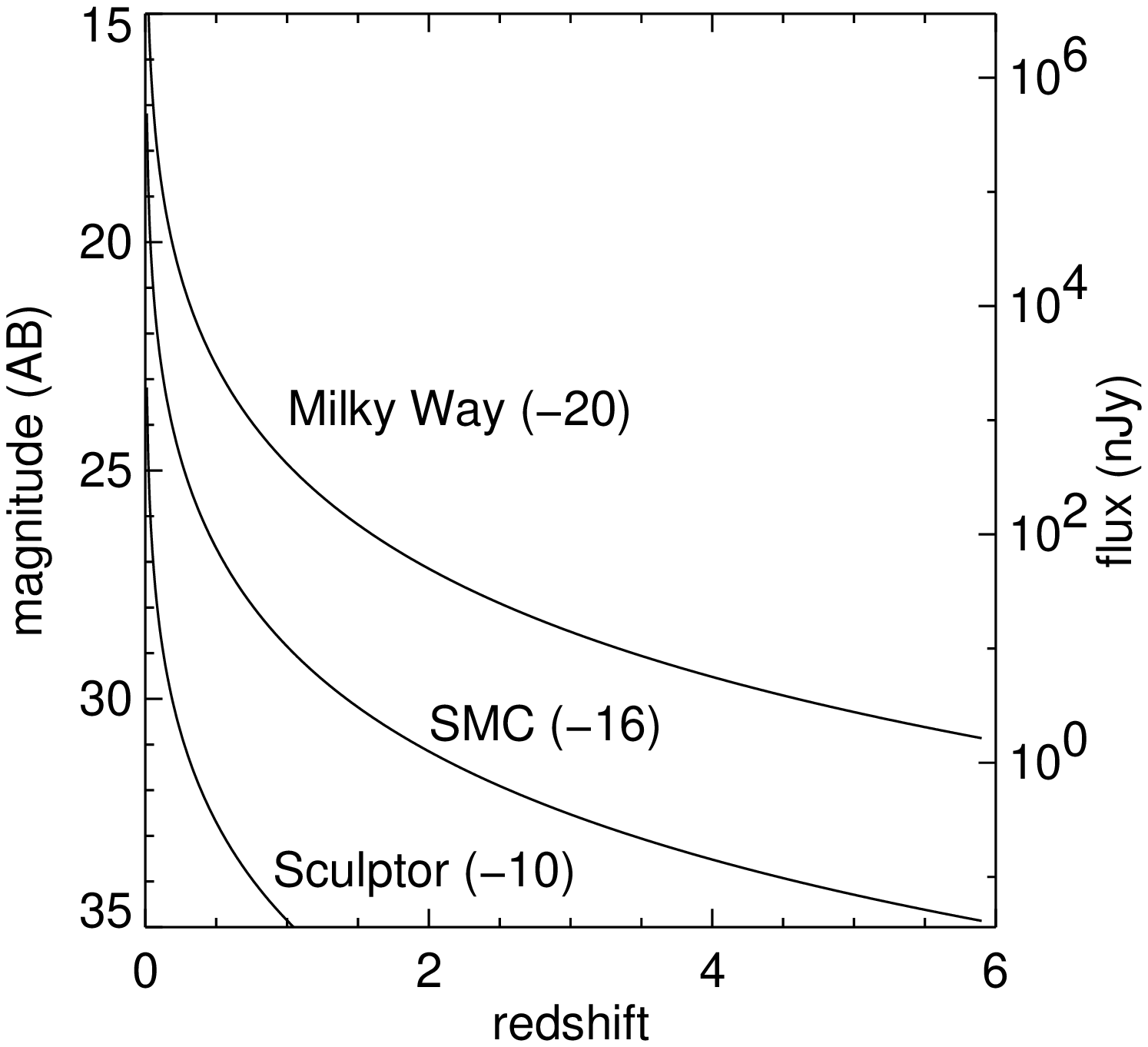}{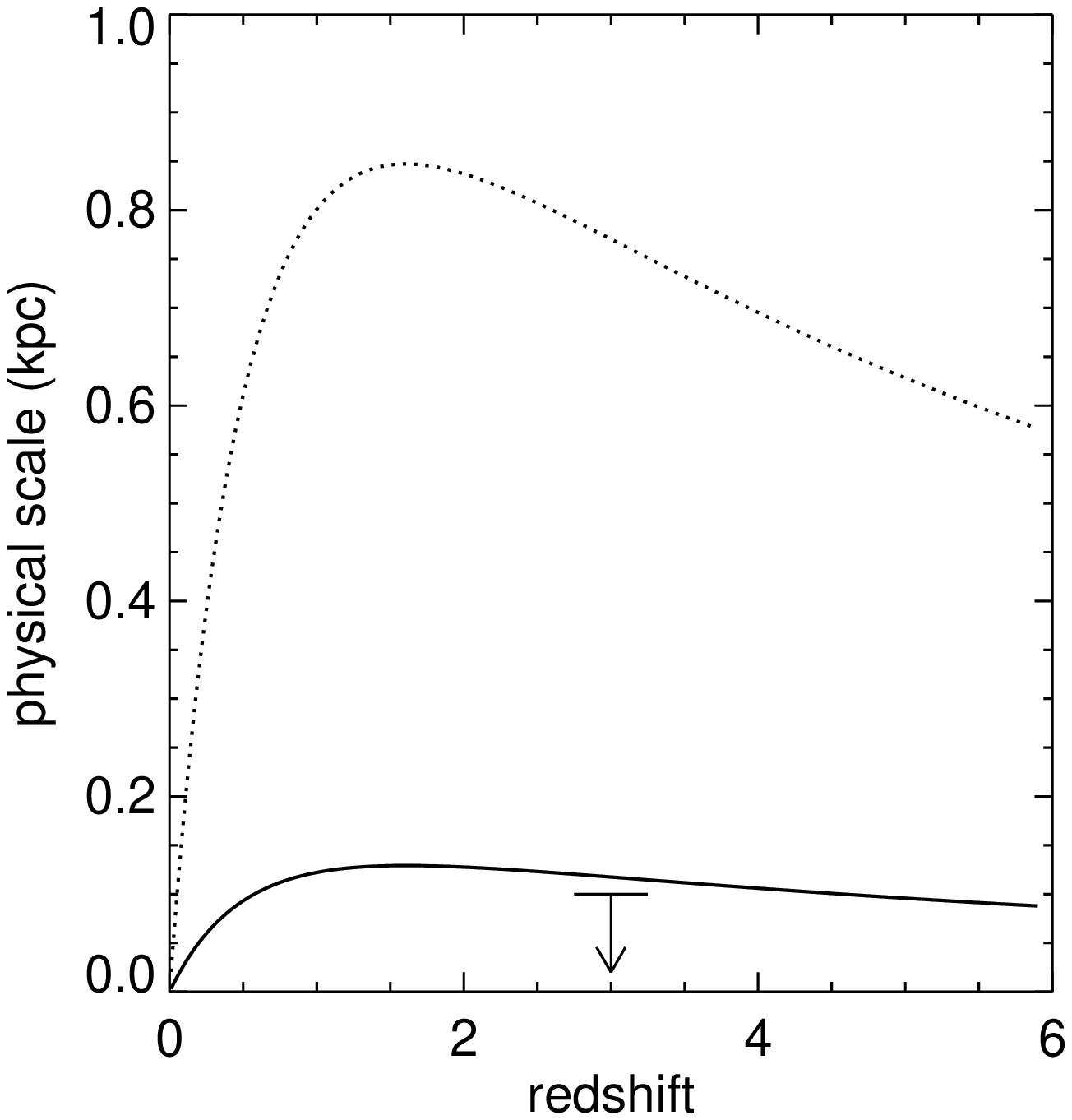}
\caption{\small {\it Left}: The apparent magnitude or flux of several
representative objects as a function of redshift. {\it Right}: Physical
scale corresponding to the approximate angular resolution of the Wide Field
Camera on HST (top line) and a diffraction limited 10m NHST (bottom
line), both at $V_{606}$. The horizontal line shows the approximate
upper limit on the required resolution necessary to study the internal
kinematics of galaxies at $z\sim3$ (see text).
\label{fig:flux}
}
\end{figure}

Quantitatively, what would the proposed parameters of NHST allow us to
do? The following calculations were done assuming a flat geometry,
$\Omega_{m}=0.3$, $\Omega_{\Lambda}=0.7$, and $H_0=70$
km~s$^{-1}$ Mpc$^{-1}$. 
Figure~\ref{fig:flux} (left) shows the apparent magnitude
(AB system) or flux of several representative objects placed at
different redshifts (no k-correction or evolution). The top line
represents a large spiral galaxy similar to the Milky Way ($M_B=-20$),
the middle line represents a typical dIrr galaxy like the SMC
($M_B=-16$), and the bottom line represents a very faint dwarf galaxy
such as Sculptor ($M_B=-10$). Assuming a flux limit on the order of a
0.5--1 nJy, which should be easily obtainable with NHST, we see that
luminous $L \simeq L_*$ galaxies will be easily visible out to
$z\sim6$. Even much fainter objects like the SMC (about three
magnitudes below $L_*$) will be visible to $z\sim2$. Objects like the
very faintest dwarf galaxies known in the Local Group (9--10
magnitudes below $L_*$) will be very difficult to see at truly
impressive redshifts --- but they should be visible out to redshifts
of $z\sim 0.5$--1, while currently these extremely tiny objects can
only be studied very locally (mostly within the Local
Group). Of course, surface brightness dimming will make these objects
even more difficult to detect at large distances.

The right panel of Figure~\ref{fig:flux} shows the physical scale
corresponding to the angular resolution of a diffraction limited 10\,m
telescope in the V-band (bottom curve). For comparison, the top curve
shows the same quantity for the approximate angular resolution of the
Wide Field Camera on HST (about 0.1 arcsec). This function is nearly flat at
$z \ga 1$, so that physical scales on the order of 0.1 kpc would be
resolved at all redshifts of interest ($z\la 6$). In the following
sections, I discuss some science goals that require this kind of
resolution.

\section{Dwarf Galaxies in the Nearby Universe}
Dwarf galaxies have long presented a particular challenge to CDM. Most
people are now familiar with the problem that CDM models tend to
produce steep luminosity functions and too many low-luminosity
galaxies (e.g., White \& Frenk 1991; Kauffmann, White, \& Guiderdoni
1993). It is usually assumed that this can be cured by invoking
supernova (SN) feedback, but most models for SN feedback are currently
quite ad hoc, and it is important to obtain better direct
observational constraints on this process. This is one of the reasons
that the study of nearby dwarf galaxies is so important.

Another form of this problem that has received a lot of attention
recently is known as the ``substructure problem''; virialized halos in
high-resolution CDM simulations have large amounts of surviving
substructure, from previous generations of dark matter halos
that have merged to form a larger halo (Klypin et al. 1999; Moore et
al. 1999). There are several hundreds of bound substructures with $V_c
\ga 20$ km~s$^{-1}$ within Local Group sized halos in the simulations, at
least an order of magnitude larger than the number of dwarf galaxies
observed in the Local Group (at most around 40).

To solve these problems within the context of standard CDM, clearly we
must invoke some mechanism to either destroy these small halos or to
prevent them from collecting gas and from forming stars. Supernova
feedback may help, but would have to be strongly differential, much
more effective in small mass objects than in large mass ones, in order
to solve this problem without making star formation too inefficient in
larger mass galaxies. Another possibility is that the presence of a
photoionizing background could suppress gas infall and cooling in very
small halos ($V_c \la 30$--50 km~s$^{-1}$). A number of recent studies using
semi-analytic methods (Somerville 2002; Benson et al. 2002; Bullock,
Kravtsov \& Weinberg 2001) concur in the conclusion that if this
effect (sometimes known as ``squelching'') is as efficient as
suggested by recent hydrodynamic simulations, then it can probably
cure the substructure problem in the Local Group.

If the ``squelching'' picture is correct, there are a number of
observational consequences that should be tested. There have been
several recent proposals for ways to probe the substructure mass
function in other galaxies using gravitational lensing (e.g., Metcalf
\& Madau 2001; Moustakas \& Metcalf 2002). We might expect that an
additional consequence of this picture would be that some small
galaxies, which formed before the Universe became reionized, would
experience a burst of early star formation but then be ``squelched''
--- these objects should then have uniformly old stellar
populations. This seems to be in conflict with the star-formation
histories derived for dwarf galaxies in the Local Group --- however,
theory predicts that the halos that can collapse before reionization
should be predominantly found in cluster environments (Tully et
al. 2002). Another implication of the squelching picture is that as
one goes down the mass function of dark matter halos, the
mass-to-light ratio of halos should greatly increase.  Since
squelching is expected to be a somewhat gradual and stochastic
process, one might be able to find very faint, surviving dwarf
galaxies which could be used as dynamical tracers of the density
field. Indeed, several candidates for these ``dwarf groups'', with
mass-to-light ratios about five times larger than those of normal
groups, have been identified (Tully et al. 2002). A direct probe of
squelching could also be obtained by measuring a ``baryonic
Tully-Fisher relation'' (or the analog for spheroidals) for extreme
dwarf galaxies ($M_B \leq -10$).

\section{Structure and Kinematics of Galactic Disks}
\begin{figure}
\plottwo{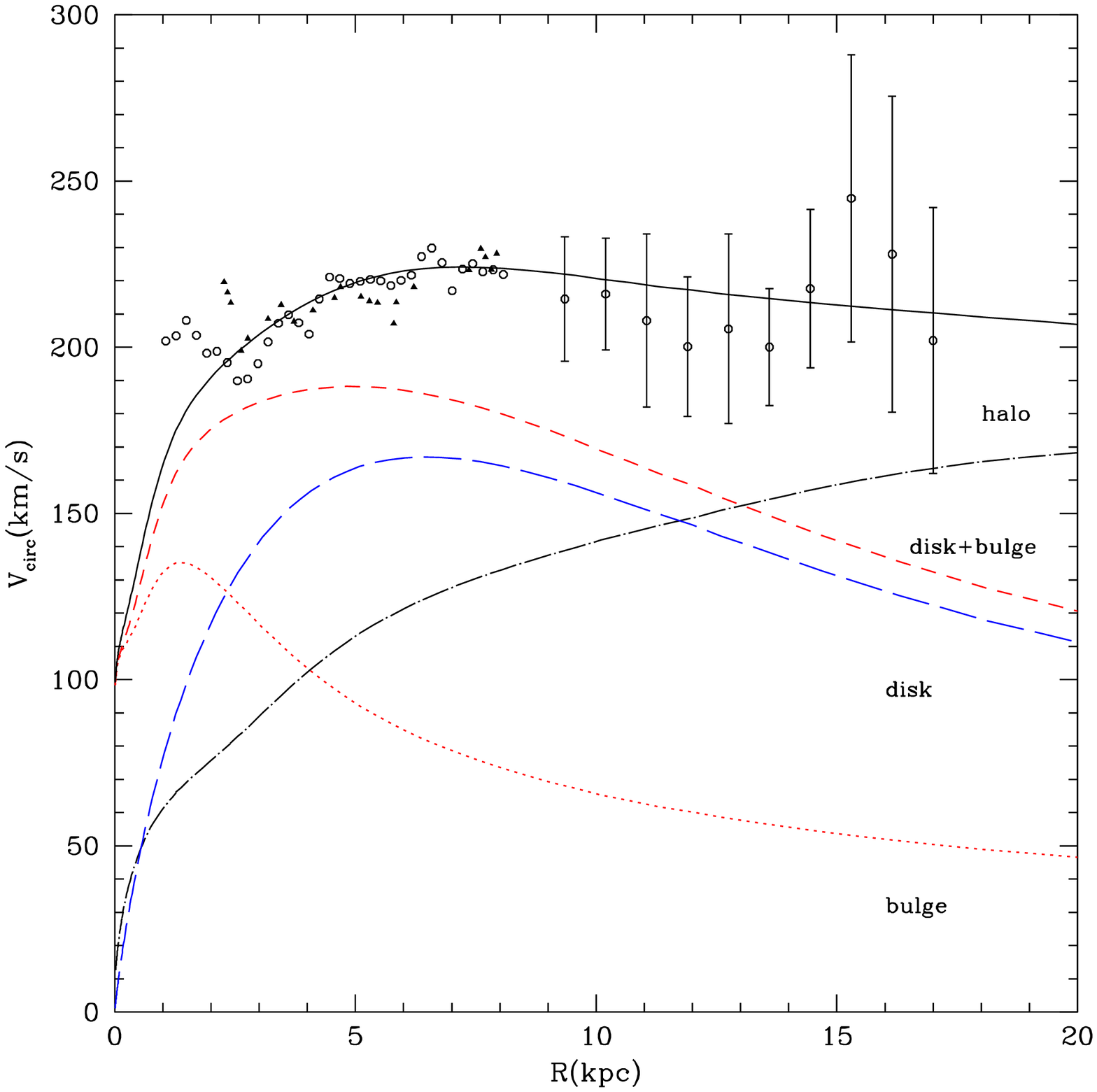}{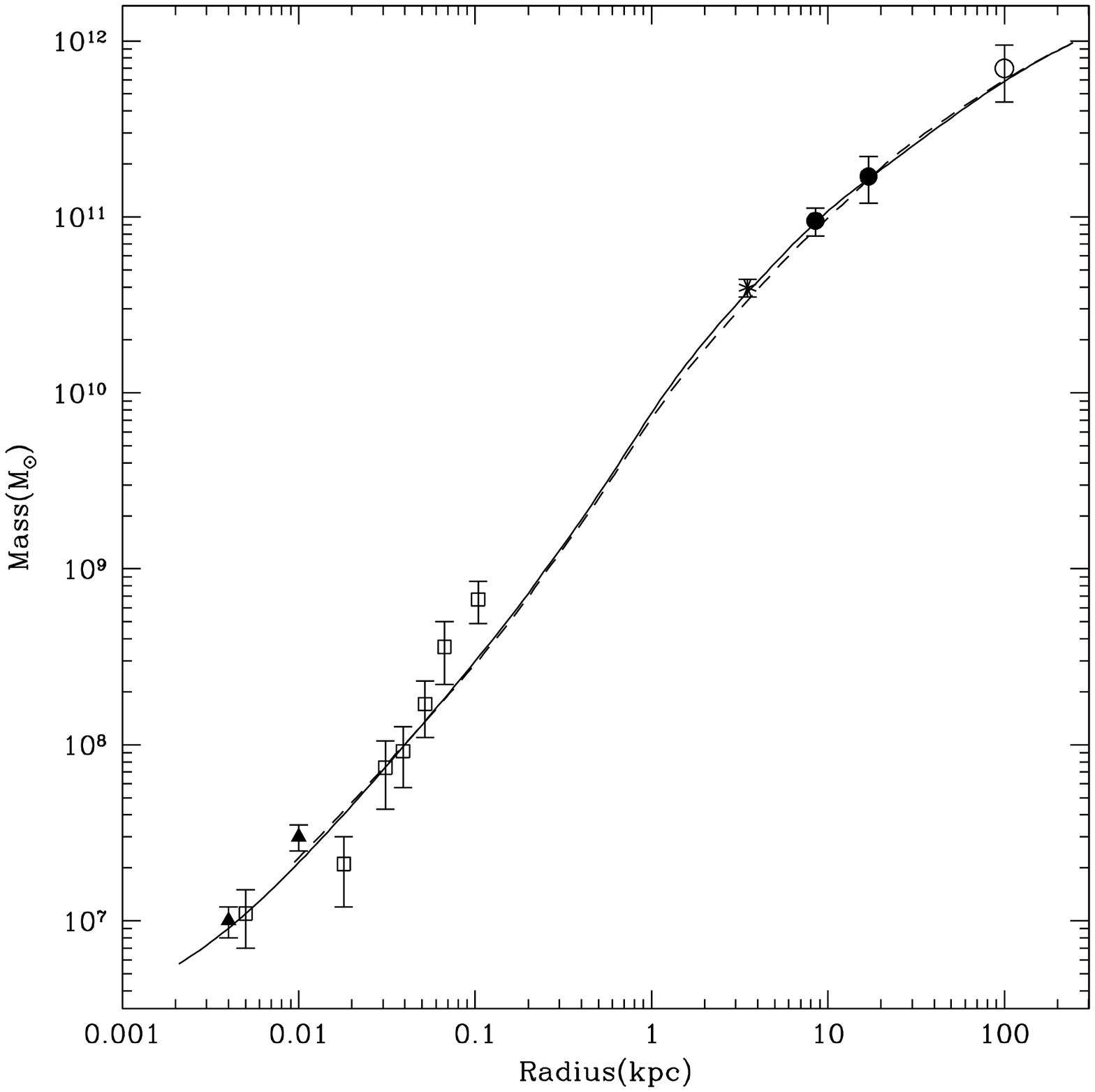}
\caption{\small {\it Left}: Rotation curve of the Milky Way. Symbols represent
observational data. Curves represent various components from the
favored model of KZS02, as indicated on the figure. In this model,
transfer of angular momentum from the baryons to the dark matter has
flattened the dark matter ``cusp'' near the center of the halo.
{\it Right}: Mass profile of the Milky Way, from pc to $\sim100$ kpc
scales. Symbols represent observational data from stellar kinematics
(inner points), H\,{\sc i} kinematics (intermediate points) and satellite
velocities (outer point). Observational references are given in
KZS02, from which both figures are reproduced.
\label{fig:MW}
}
\end{figure}

Another potentially serious problem for CDM-based models is the set
of issues surrounding the internal structure of disk galaxies.  It is
now well-known that the density profiles of dark matter halos formed
in dissipationless CDM simulations are ``cuspy'' -- i.e., they rise
fairly steeply in the center, with a slope of about $r^{-1}$ (Navarro,
Frenk, \& White 1997 (NFW); Bullock et al. 2001; Power et al. 2002).
It is difficult to reconcile the observed rotation curves of some
dwarf and low surface brightness galaxies with these NFW halos, which
seem to predict too much dark matter in the central part, leading to
rotation curves that rise too abruptly. There is considerable
disagreement, however, about whether there really is a problem or
whether the discrepancies can be explained by selection effects or
observational errors. All of the problematic objects are from a
subclass of extremely late type, dwarf and/or low surface brightness
(LSB) galaxies. Perhaps this population can be accounted for if they
are formed within halos drawn from an extreme tail of low
concentrations or flat inner profiles. In order to lay this debate to
rest, we need high resolution kinematics for a statistically complete
sample of galaxies, extending to very low luminosities and surface
brightnesses. The existing samples are sparse because these tiny,
faint objects are so difficult to detect, and even more difficult to
obtain good spectra for. The resolution, sensitivity, and high S/N
that could be obtained with NHST could provide the necessary
breakthrough to determine whether this is indeed a ``crisis'' for CDM.

If there really is a problem, perhaps it can be solved by appealing to
baryonic processes. It is generally argued that this is unlikely for
the objects in question, which are clearly dark matter dominated in
the center. To go this route, one must appeal to an event that occurred
at an early stage, leaving no directly observable trace. For
example, cusps could be destroyed by an early, intense burst of star
formation which simultaneously drove all the gas out of the halo,
perhaps combined with subsequent tidal heating (Gnedin \& Zhao 2001;
Dekel \& Devor 2002). Weinberg \& Katz (2002) proposed that a large
and massive rotating bar in the high-redshift progenitors of present
day galaxies could destroy the inner dark matter cusps. These
solutions all involve events which should produce rather dramatic
observational signatures but which must occur at high enough redshift
that they are safe (for the moment) from direct confrontation with
observations. Perhaps the best way to test these ideas is to look
directly at the high redshifts where these processes are assumed to
occur. NHST would be very useful for studying the internal structure
and dynamics of the progenitors of present day galaxies, which may
provide constraints on these kinds of scenarios.

It has also been suggested that there may be too much dark matter in
the centers of luminous galaxies, like our own Milky Way (e.g., Binney
\& Evans 2001). Klypin, Zhao, \& Somerville (2002; KZS02) found,
however, that if there is moderate transfer of angular momentum from
the baryons to the dark matter (about a factor of two), the detailed
inner structure and kinematics of the Milky Way and M31 can be
understood within the context of NFW halos (see
Figure~\ref{fig:MW}). The analytic solution of KZS02 is reminiscent of
the process seen by Weinberg \& Katz (2002) in their simulations.

One of the main uncertainties in interpreting galaxy rotation curve
data is due to the difficulty in connecting the virial mass of the
dark matter halo (typically, a few hundred kpc for bright galaxies)
with the observed rotation curves, which probe only about the inner
tenth of the halo. Especially for luminous galaxies, the gravitational
force of the baryons is expected to significantly modify the profile
in the central parts, where the baryons comprise about half of the
mass. For the unique case of the Milky Way, we can reconstruct the
mass profile over a huge range of scales (see Figure~\ref{fig:MW}) by
using different dynamical tracers: stellar kinematics on small scales
(pc), the H\,{\sc i} rotation curve on intermediate scales (kpc), and
satellite velocities on large scales (100 kpc). As shown in KZS02,
these combined data produce very strong constraints on the
distribution of baryons and dark matter in the Milky Way. NHST could
be used to obtain very high resolution kinematic data for the central
parts of galaxies, and kinematic tracers of the outermost reaches of
the halos of galaxies such as satellite galaxies, planetary nebulae,
or globular clusters. These data could be combined with CO rotation
curves from the Atacama Large Millimeter Array (ALMA)
to obtain very tight constraints on the distribution
of mass out to the virial radii of galaxies. This will be invaluable
in understanding the complex process of disk formation and what
determines the surface density profile and rotation curve of spiral
galaxies. Similar exercises may be done for early type galaxies.

\section{High Redshift Galaxies: What Are They?}
\begin{figure}
\plottwo{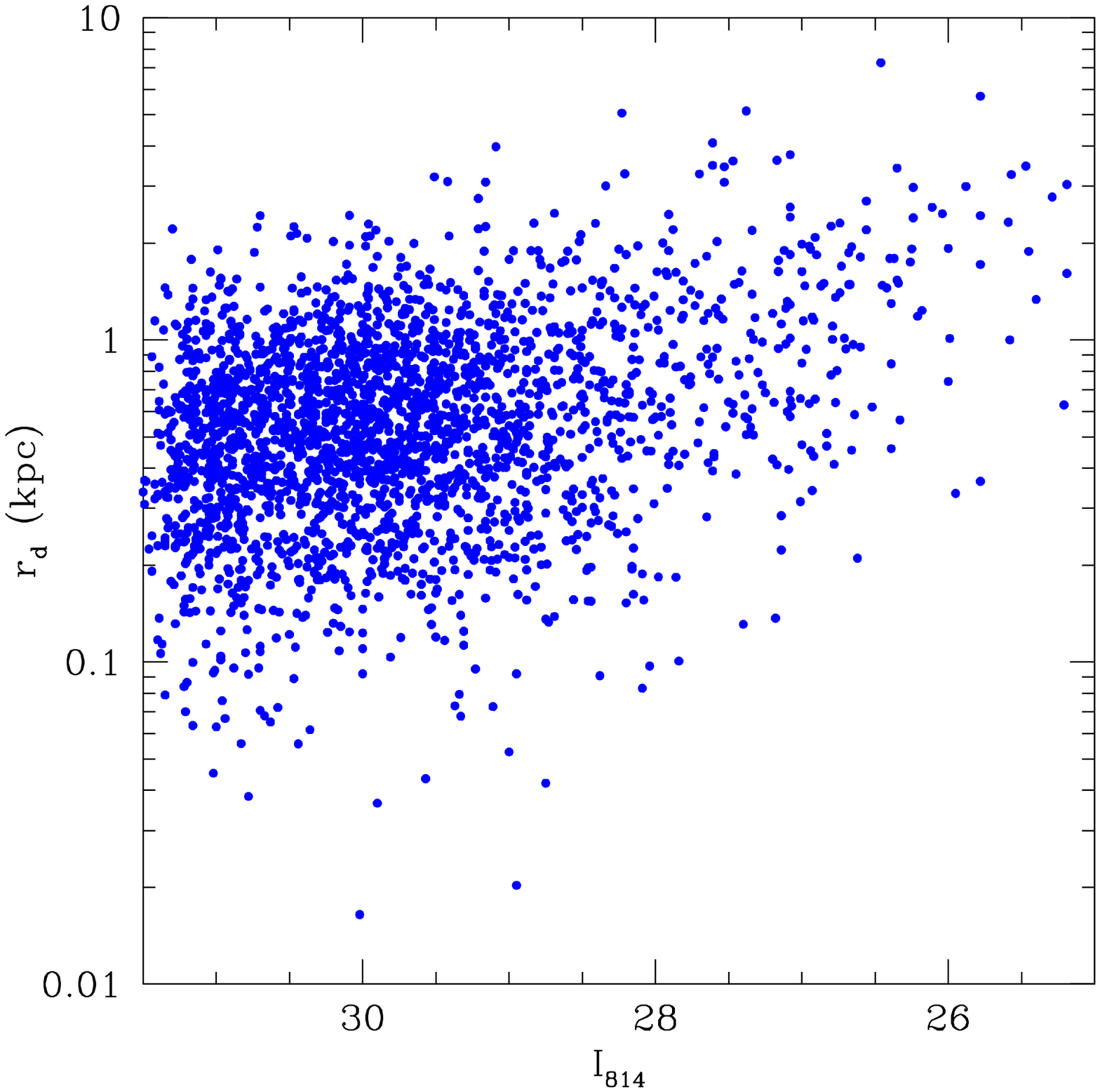}{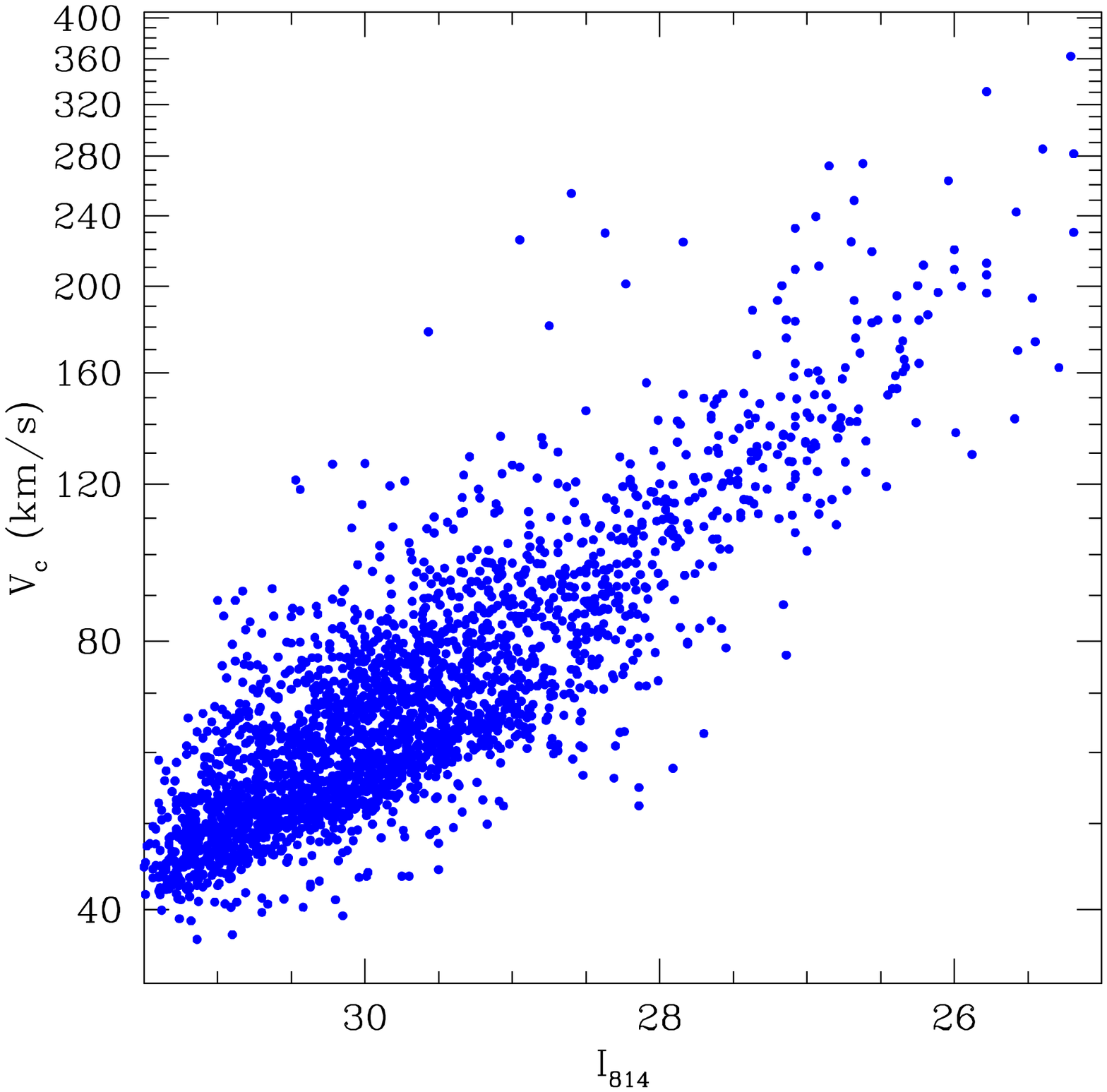}
\caption{\small {\it Left}: exponential disk scale length versus apparent magnitude
(observed frame) for model disk galaxies at $z\sim3$. {\it Right}:
Tully-Fisher relation for model disk galaxies at $z\sim3$.
\label{fig:struc}
}
\end{figure}
We now know of the existence of well over a thousand galaxies at
$z\ga2$. While this is an enormous achievement, in many respects we
still do not know what these objects are. Are they disks or
spheroids? Are their star-formation rates so much higher than
present-day galaxies because they have been brightened by collisions,
or because their disks have higher gas surface densities? Do they
follow the same fundamental scaling relations as nearby galaxies? What
objects in the lower redshift Universe are their descendants?
Arguably, we will not really know what these objects are until
we image them at considerably higher resolution, and ideally, obtain
spatially resolved kinematic data for them. Combined with information
on the gas distribution from ALMA, this could allow us to test whether
empirical star-formation laws like the Kennicutt law (1989; 1998) are
universal.

What kind of resolution would this require? Figure~\ref{fig:struc}
shows the predictions of a CDM-based semi-analytic model of galaxy
formation for the structural properties of galaxies at $z\sim3$.  This
model is similar to those described in Somerville \& Primack (1999)
and Somerville, Primack, \& Faber (2001; SPF), but includes a more
detailed model of disk formation within cosmological ``NFW'' halos,
including the effects of adiabatic contraction of the baryons using an
approach similar to the one described in Mo, Mao, \& White (1998). The
left panel of Figure~\ref{fig:struc} shows the disk scale length (in
kpc) versus the $I_{814}$ magnitude (AB). A couple of things are worth
noting. While the predicted sizes of the brighter objects are
consistent with those measured for Lyman break galaxies in the Hubble
Deep Field (see SPF), the models predict that these objects are only
the tip of the iceberg; there is a large population of objects too
small and too faint for HST to resolve. The right panel shows the
Tully-Fisher relation for the same model galaxies at $z=3$ (note that
the quantity plotted is the observed frame I-band, corresponding to
rest far-ultraviolet, so it is not surprising that there is a large scatter in
this relation). Note the break in this relation at around $I_{814} >
28$, which is caused by the inclusion of ``squelching'' in the models.

\section{Summary}
I have suggested several applications of the proposed NHST 8-10\,m
ultraviolet/optical space telescope to the study of galaxies. A concise summary
of these suggestions follows:
\begin{itemize}
\item Study stellar populations of dwarf galaxies (constrain their 
star-formation histories, and test ``squelching'' scenario).
\item Look for ``dwarf groups'' with high mass-to-light ratios (the
tail of the ``squelched'' population).
\item Collect high resolution kinematic data for a complete sample of
nearby galaxies, including dwarf and LSB galaxies (assess ``cusp''
crisis).
\item Constrain the dark matter around galaxies on scales of $\sim100$
kpc using satellite galaxies, planetary nebulae, and/or globular
clusters as kinematic tracers (understand disk and spheroid formation).
\item Obtain high resolution imaging and spatially resolved kinematics
of $z\ga2$ galaxies.
\item Obtain star-formation indicators and gas surface densities for a
broad range of environments and redshifts, to study the universality
of empirical star-formation scaling laws.
\end{itemize}

\end{document}